\documentclass[aps,prl,twocolumn,superscriptaddress]{revtex4-1}

\usepackage{amsmath,amssymb}
\usepackage{graphicx}
\usepackage{tikz}
\usepackage{amsmath}          
\usepackage[dvipsnames]{xcolor} 
\usepackage{verbatim}
\usetikzlibrary{calc,positioning}

\tikzset{
  external/.style={draw=black, thick},
  internal/.style={draw=black, thick},
  vertex/.style={circle, fill=black, inner sep=1.7pt},
  momentum/.style={font=\small},
  variable1/.style={font=\small}
}

\begin{document}

\title{Cosmological Polytopes and the Emergence of Time(s)}

\author{Nima Arkani-Hamed}
\email{arkani@ias.edu}
\affiliation{School of Natural Sciences, Institute for Advanced Study, Princeton, New Jersey 08540, USA}

\author{Carolina Figueiredo}
\email{cfigueiredo@fas.harvard.edu}
\affiliation{Society of Fellows, Harvard University, Cambridge, Massachusetts 02138, USA}

\author{Ross Glew}
\email{r.glew@ias.edu}
\affiliation{School of Natural Sciences, Institute for Advanced Study, Princeton, New Jersey 08540, USA}

\author{Francisco Vaz\~ao}
\email{fvvazao@ias.edu}
\affiliation{School of Natural Sciences, Institute for Advanced Study, Princeton, New Jersey 08540, USA}
\affiliation{University of Pennsylvania, Philadelphia, Pennsylvania 19104, USA}

\begin{abstract}
The late-time cosmological wavefunction is usually computed perturbatively via explicit time evolution, with the contribution from any graph given in terms of time integrals over the bulk vertex interactions. A radically different picture for the wavefunction of simple toy scalar models is given by the canonical form of a positive geometry -- the ``cosmological polytope". But how can the conventional picture of time evolution be seen to emerge from the positive geometry? In this letter we show that ``time integrals" arise in the most direct possible way, from the simplest representation of the canonical form as a dual volume. This is conveniently computed by the Laplace-transform over the dual cone of the cosmological polytope, expressing the wavefunction as an integral over both edge- and vertex-variables of the graph. Integrating over the edge variables precisely recovers the usual bulk time integral representation. Dually, integrating out the vertex variables gives rise to a new representation with a different notion of ``time" associated with edges, connecting to the formulation of wavefunctions in terms of graph associahedra.    
\end{abstract}

\maketitle

\section{Introduction}
\label{sec:introduction}

The past fifteen years have seen the emergence of a new way of conceptualizing some of the most basic observables in fundamental physics, from scattering amplitudes to flat-space and cosmological wavefunctions. The usual formalism of summing over spacetime trajectories and evolution in Hilbert space is replaced by a radically different picture, where the observables are associated with canonical forms of positive geometries living directly in kinematic space. These geometries have their own, autonomous definition, determined by simple, combinatorial questions, from which the principles of locality and unitarity emerge as outputs rather than inputs. 

In the context of wavefunctions and cosmology, much of this work has centered around the simple toy model of conformally coupled scalars in general FRW cosmologies, whose wavefunction coefficients are very simply related to the corresponding wavefunctions in flat space. Of course, these wavefunctions have a direct, textbook ``time-integral" representation using Feynman diagrams. For a fixed graph, these objects were interpreted as the canonical form of ``cosmological polytopes", in the first instance of the connection between wavefunctions and positive geometry. More recently, the wavefunctions were given a different interpretation, in terms of canonical forms of ``graph associahedra".  

\begin{figure}[]
    \centering
    \includegraphics[width=\linewidth]{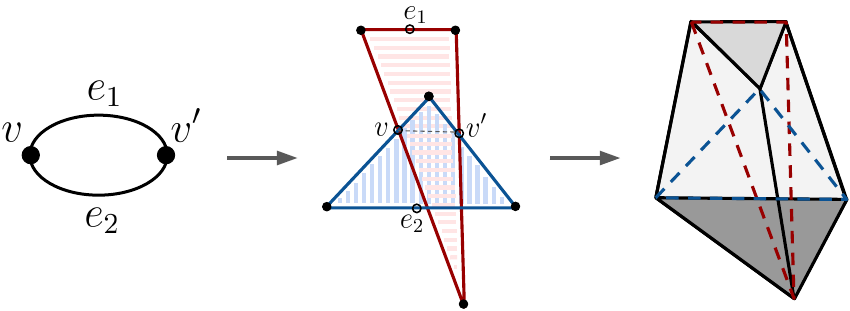}
    \caption{For any graph $G$, associate a  triangle with every edge $e$ of $G$, with the midpoints of the triangle's edges marked with the edge $e$ and the two vertices $v, v^\prime$ meeting $e$. If two edges $e$ and $\hat{e}$ share vertex $v$, then the respective triangles intersect each other at $v$. The convex hull of all these triangles is the cosmological polytope of $G$. }
    \label{fig:CosmoPoly}
\end{figure}

Nonetheless, the cosmological polytope has always seemed to be the most fundamental of these objects. The definition of a cosmological polytope for a given graph is amazingly simple: each internal edge $e$ of the graph is associated with a triangle, $T_e$. Two of the midpoints of the sides of $T_e$ are associated with the two vertices that are endpoints of $e$, and the third midpoint of $T_e$ is associated with $e$ itself. We then glue these triangles together by gluing the midpoints associated with the same vertices of the graph. Taking the convex hull of this collection of intersecting triangles produces the cosmological polytope (see Fig.\ref{fig:CosmoPoly}). There is not a shred of reference to any notion of ``time evolution" in this picture, and yet the polytope's canonical form computes the same wavefunction coefficients as we conventionally obtain from the time integrations of Feynman diagrams. 
Despite all of these developments, these ideas have yet to address the most obvious and deepest open question: how can we understand the ``emergence of time" from combinatorial geometry? While we do have two different ways of computing the same wavefunction: the textbook one in terms of time integrals, and the positive-geometric one via canonical forms; this is not the same as seeing the emergence of time, since the latter can be defined without any time interpretation whatsoever. Is there a way of naturally {\it discovering the time evolution representation}, starting from the computation of the canonical form for positive geometries?  

In this letter, we show that the cosmological polytope gives an astonishingly direct answer to this question: the time integral representation falls into our lap with no contrivance of any sort if we start with the best-known and most fundamental representation for the canonical function of any polytope, expressed as a Laplace transform over the dual of its associated cone. In coordinates adapted to the incidence structure of the graph , this takes the particularly simple form
\[
\scalebox{0.92}{$\displaystyle
\begin{aligned}
\Psi_G &=
\int_0^\infty
\prod_{v\in V}
\left(d\tau_v\,e^{-p_v\tau_v}\right)
\prod_{e\in E}
\left(d\tau_e\,e^{-q_e\tau_e}\right)
\prod_{v\sim e}\Theta(\tau_v-\tau_e)
\end{aligned}$}
\]
where $(p,q)$ are related to the usual energy variables. The remarkable feature of this representation is that both the vertex and edge integrals can be performed immediately.

Performing the edge integrals yields precisely the bulk-to-bulk propagators in the original energy variables, recovering the Feynman representation as an integral over vertex-interaction times. It is hard to imagine a more direct manifestation of the \emph{emergence of time} from combinatorial geometry: the time-ordering of Feynman diagrams follows immediately upon performing a subset of the integrations in the Laplace transform over the dual cone of the cosmological polytope!

Remarkably, the vertex integrals can be performed just as simply. This gives a new integral representation of $\Psi_G$ entirely for free, with a different, dual encoding of bulk-time evolution: the integration variables are instead associated with the edges of the graph and are coupled only when those edges meet at a common vertex, giving precisely an integral representation for the canonical form of graph associahedra! 

Beautifully, then, we have obtained two vertex--edge dual interpretations of bulk-time evolution, each with its own set of Feynman rules, both arising naturally from the geometry of the cosmological polytope.

We further show how two recursion relations familiar from the literature arise from the mechanism of integrating out the smallest time variable. In the vertex representation, the smallest interaction time leads to a vertex-deletion recursion for the connected-propagator contribution. In the edge representation, the smallest edge variable instead leads to an edge-deletion recursion for the full wavefunction coefficient, recovering the familiar \emph{old-fashioned perturbation theory} representation obtained by solving the Schr\"odinger equation perturbatively. Iterating these recursions recovers two related but distinct notions of graph tubings, which have previously appeared in the literature to expand the wavefunction coefficients.

\section{The Wavefunction}
\label{sec:laplace}

\begin{figure}[]
    \centering
    \includegraphics[scale=1.1]{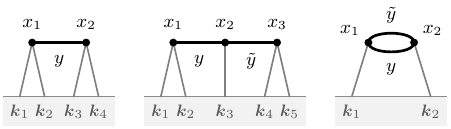}
    \caption{Graph contributions to the wavefunction coefficient.}
    \label{figs:feyn}
\end{figure}

We consider the wavefunction for a massless scalar with polynomial interactions in flat space \cite{CosmoPolys,Cosmohedron,Anninos:2014lwa,Arkani-Hamed:2018kmz,
DiffEq_CosmoCorr,KinFlow,He:2024olr,Baumann:2024mvm,De:2024zic,Capuano:2025ehm,
GeometryCosmoCorr,Benincasa:2024lxe,Goodhew:2020hob,CosmoReview,IRSub,MelvillePajer,PajerUnitarityLocality,GoodhewCutCosmoCorr,CosmoSMatrix,Sachs2,CosmoTreeTheo,Meltzer,Bittermann,Benincasa:2024ptf,CespedesScott,Sachs3,Thavanesan:2025kyc,Goodhew:2024eup,Forcey:2025voc,Glew:2025ypb,HiddenZerosWF,Ardila-Mantilla:2026cbo,Benincasa:2025uie,Glew:2025otn,Glew:2025ugf,Raman:2025tsg,Xianyu:2025lbk,Benincasa:2019vqr,Fan:2024iek,Figueiredo:2025daa,Figueiredo:2026tcj,Zhu:2026xil,Glew:2025arc,Glew:2026von,Hang:2026men,Hang:2024xas},
\begin{equation}
 \Psi[\Phi]=\mathcal N
 \int_{\varphi(\tau=0,\mathbf x)=\Phi(\mathbf x)}
 \mathcal D\varphi\,e^{-S_E[\varphi]},
 \label{eq:vacuum-wavefunctional}
\end{equation}
here $S_E$ is the Euclidean action \cite{Higuchi:2010xt}, obtained by Wick rotating Lorentzian time $\eta\in(-\infty,0]$ according to $\eta=i\tau$, with $\tau\in[0,\infty)$. The vacuum prescription selects solutions decaying as $\tau\to\infty$.

Expanding the path integral, we obtain a perturbative sum of Feynman graphs, see fig.~\ref{figs:feyn}, where each vertex carries a time integral, each internal edge a bulk-to-bulk propagator, and each external edge a bulk-to-boundary propagator. For a connected graph $G=(V,E)$, the corresponding time integral is:
\begin{align}
 \Psi_G=\int_0^\infty
 \prod_{v\in V}\left(\mathrm d\tau_v\,e^{-x_v\tau_v}\right)
\prod_{e \in E}G_{\rm BB}(\tau_v,\tau_{v^\prime};y_e),
 \label{eq:intro-graph-integral}
\end{align}
Here $x_v$ is the sum of external energies entering vertex $v$, $y_e$ is the magnitude of the momentum through edge $e$, and
\begin{equation}
 G_{BB}(\tau,\tau';y)
 =\frac{e^{-y|\tau-\tau'|}-e^{-y(\tau+\tau')}}{2y},
 \label{eq:intro-propagator}
\end{equation}
where the second term ensures that the propagator vanishes at the boundary. Upon time-integration, we obtain a rational function of the vertex and edge energies, $x_v$ and $y_e$, where the singularities are located where sums of energies entering subgraphs of $G$ go to zero. For example, one singularity of the five-point graph in Fig.~\ref{figs:feyn} is associated to the subgraph $\mathcal{E}_{12}=\vcenter{\hbox{\includegraphics[scale=0.6]{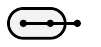}}} = x_1 + x_2+ \tilde{y} =0$. 

\section{Cosmological Polytopes}
In \cite{CosmoPolys}, it was shown that $\Psi_G$ can also be interpreted as the canonical form of a polytope -- the \textit{cosmological polytope}. Given a connected graph $G$,  with $|V|$ vertices and $|E|$ edges, we can build the respective cosmological polytope by assigning a triangle to each edge, identifying their midpoints according to the incidence relations of the graph, and taking the convex hull of the resulting collection of intersecting triangles.  

Concretely, let $n=|V|+|E|$, and consider the vector space $\mathbb{R}^n$, with basis vectors ${\bf X}_v$ for each $v \in V$, and ${\bf Y}_e$ for each $e \in E$. A general vector in this space can be written as 
\begin{align}
\mathcal Y=\sum_vx_v {\bf X}_v+\sum_ey_e {\bf Y}_e.
\end{align}
The cone over the cosmological polytope, $ \mathcal C_G$, is the positive span of the following $3|E|$ vectors
\begin{equation}
 \mathcal C_G=\operatorname{span}_{\mathbb{R}^+}\left\{
 \begin{aligned}
 &+{\bf X}_v+{\bf X}_{v^\prime}-{\bf Y}_e,\\
 &+{\bf X}_v-{\bf X}_{v^\prime}+{\bf Y}_e,\\
 &-{\bf X}_v+{\bf X}_{v^\prime}+{\bf Y}_e
 \end{aligned}
 \ \middle|\ 
 \begin{gathered}
    \raisebox{-4pt}{
    \begin{tikzpicture}[scale=0.6,font=\small,
    line cap=round,
    line width=1pt,
    offshell/.style={
        draw=Black,
        line width=1.5pt
    },
    onshell/.style={
        draw=gray,
        line width=0.9pt
    },
]
    \draw[onshell] (-0.8, 0.6) -- (0,0);
    \draw[onshell] (-0.4,0.6) -- (0,0);
    \draw[offshell] (0,0) -- (1,0);
    \draw[onshell] (1, 0) -- (1.4,0.6);
    \draw[onshell] (1,0) -- (1.8,0.6);
    \node at (0.5, 0.3) {$e$};
    \node at (-0.3, -0.2) {$v$};
    \node at (1.4, -0.2) {$v^\prime$};
    \node at (0.5, -0.9) {$\subset G$};

    \filldraw[black] (0,0) circle (2pt);
    \filldraw[black] (1,0) circle (2pt);
\end{tikzpicture}}
    \end{gathered} \right\}.
 \label{eq:cosmological-polytope}
\end{equation}
The facets of $\mathcal{C}_G$ are labeled by connected subgraphs or tubes of $G$, which correspond to the singularities of $\Psi_G$. Given a tube $H \subseteq G$, the associated facet lies on
$\mathcal E_H=0$, with 
\begin{equation}
 \mathcal E_H
 =\sum_{v\in V_H}x_v+\sum_{e\in \partial H}y_e,
 \label{eq:subgraph-energy}
\end{equation}
which is simply the total energy entering the tube, once we interpret $x_v$ and $y_e$ as the energy variables. Here \(\partial H\) consists of the edges incident to but not contained in \(H\), counted once for each endpoint in \(H\). The facet associated with the entire graph is referred to as the \textit{scattering facet}, and corresponds to the \emph{total energy} singularity, $\mathcal{E}_G \equiv \mathcal{E}_{\rm tot} = x_1+\ldots+x_{|V|}$.

Viewed projectively, the cone $\mathcal C_G$ defines the cosmological polytope, whose canonical form is the product of the standard projective measure and a rational canonical function. This canonical function is then $\Psi_G$ and, as we review below, admits a representation as a Laplace transform over the dual cone $\mathcal C_G^\vee$. 
\\ \\
\textbf{Incidence coordinates ---} Before studying the Laplace representation of the wavefunction, we introduce the following useful change of coordinates
\begin{equation}
 p_v=x_v+\sum_{e\ni v}y_e,\qquad q_e=-2y_e.
 \label{eq:perimeter-variables}
\end{equation}
where $p_v$ equals the sum of energies entering a single vertex, $v$. This coordinate change is accompanied by the following change of
basis vectors
\begin{align}
{\bf P}_v = {\bf X}_v, \qquad 2 {\bf Q}_e = {\bf X}_{v}-{\bf Y}_e + {\bf X}_{v^\prime}.
\end{align}
In these new coordinates, up to an overall factor of two, the generators of the cone over the cosmological polytope become
\begin{align}
\mathcal{C}_G = \operatorname{span}_{\mathbb{R}^+
}\{{\bf P}_v -{\bf Q}_e, {\bf Q}_e,{\bf P}_{v^\prime} -{\bf Q}_e : e  \in E \}.
\end{align}
The facet form $\mathcal{E}_H$ for each connected subgraph is now given by
\begin{align}
\mathcal{E}_H = \sum_{v \in V_H} p_v+\sum_{e \in E_H} q_e.
\end{align}
Here $V_H$ and $E_H$ denote respectively the vertices and edges inside $H$. For example, the total energy is now expressed as $\mathcal{E}_{\rm tot} = \sum_{v \in V} p_v + \sum_{e \in E} q_e$.
\\ \\
\textbf{Laplace transform --- }The canonical function of the cosmological polytope admits a representation as the Laplace transform over the cone dual to $\mathcal C_G$. Knowing the generators of this cone, its dual is immediately expressed in coordinates $(\tau_v,\tau_e)$ dual to $(p_v,q_e)$ as
\begin{align}
\mathcal{C}_G^\vee = \{ (\tau_v,\tau_e) \in (\mathbb{R}^n)^* \ |  & \ 0\leq \tau_e \leq \tau_v,\tau_{v'} \ \forall \ e \in E \}.
\end{align}
These inequalities make the advantage of the new coordinates explicit: fixing either the vertex or edge variables leaves independent intervals for the other. Consequently, either family of integrations in the Laplace transform can be performed independently once the other variables are held fixed. In the original dual coordinates, fixing the edge variables instead leaves coupled constraints on the vertex variables, and hence coupled vertex integrations.

 The graph wavefunction coefficient is the Laplace transform of $\mathcal{C}_G^\vee$, given by
\begin{align}
\Psi_G = \int_0^\infty d^{|V|+|E|}\tau \, &e^{ -\sum_{v} p_v \tau_v -\sum_{e} q_e \tau_e} \,  \prod_{v \sim e} \Theta_{ve},
\label{eq:laplace-perimeter}
\end{align}
where the product is over all vertex-edge incident pairs and $\Theta_{ve} \equiv\Theta(\tau_v -\tau_{e})$.

For example, for the two-site chain graph, the first graph in fig.~\ref{figs:feyn}, the integral is 
\begin{equation}
\begin{aligned}
\Psi_{\begin{gathered}
    \begin{tikzpicture}[line width=0.9,scale=0.15,line cap=round,every node/.style={font=\scriptsize}]
        \coordinate (L1) at (-1,0);
        \coordinate (R1) at (1,0);
        \draw (L1) -- (R1);
        \filldraw[color=black,fill=black] (L1) circle[radius=0.2];
        \filldraw[color=black,fill=black] (R1) circle[radius=0.2];
\end{tikzpicture}
\end{gathered}} = \int_0^\infty d \tau_1 d\tau_2 d&\tau_{12}  \,e^{-p_1\tau_1-p_2\tau_2-q_{12} \tau_{12}} \\
& \, \, \, \, \times \Theta(\tau_1-\tau_{12}) \Theta(\tau_2-\tau_{12}).
\end{aligned}
\end{equation}
Performing the integrations produces $\frac{1}{p_1p_2(p_1+p_2+q_{12})}$, which after substituting \eqref{eq:perimeter-variables} reduces to the familiar expression for the two-site chain wavefunction. 
\section{Vertex-edge duality}
\label{sec:integral-representations}
Starting from the Laplace integral \eqref{eq:laplace-perimeter}, either the vertex or edge integrations can be carried out directly. This yields two vertex--edge dual interpretations of the same graph. In the first, time variables reside on vertices and are coupled across edges by bulk-to-bulk propagators, recovering the usual bulk time integral representation of the Feynman diagram. In the second, time variables reside on edges and couple only when their edges meet at a common vertex, resulting in a dual interpretation of the same diagram.
\\ \\
\textbf{Discovering Time Evolution --- }
Grouping the theta functions of \eqref{eq:laplace-perimeter} by edge restricts each edge variable to be less than both of its endpoint vertex variables. The edge-dependent part can be integrated and directly yields the bulk-to-bulk propagator
\begin{align}
 G_{BB}(\tau_u,\tau_v;y_e)
 &=e^{-y_e(\tau_u+\tau_v)}
 \int_0^{\min(\tau_u,\tau_v)}\!\mathrm d\tau_e\,e^{2y_e\tau_e}
 \nonumber\\
 &=\frac{e^{-y_e|\tau_u-\tau_v|}
       -e^{-y_e(\tau_u+\tau_v)}}{2y_e}.
 \label{eq:bulk-propagator}
\end{align}
This organizes \eqref{eq:laplace-perimeter} immediately into the standard time integral representation
\begin{equation}
 \Psi_G=\int_0^\infty\mathrm d^{|V|}\tau\,
 e^{-\sum_vx_v\tau_v}\prod_{e=uv\in E}G_{BB}(\tau_u,\tau_v;y_e).
 \label{eq:vertex-integral}
\end{equation}
The emergence of time from the cosmological polytope is now completely manifest: the dual vertex variables $\tau_v$ acquire the interpretation of interaction times connected by bulk-to-bulk propagators. The familiar Feynman picture is therefore recovered from the geometry alone, with no additional physical input.
\\ \\
\textbf{Dual Edge Times --- }
Grouping the theta functions by vertex restricts each vertex variable to be greater than its incident edge variables. The vertex-dependent part of \eqref{eq:laplace-perimeter} can be integrated directly to give:
\begin{align}
\int_{\max_{e\ni v}\tau_e}^{\infty}d\tau_v\,e^{-p_v\tau_v}
=\frac{e^{-p_v\max_{e\ni v}\tau_e}}{p_v}.
\end{align}
Substituting into \eqref{eq:laplace-perimeter}, the wavefunction
becomes an integral over edge variables
\begin{align}
 \Psi_G =\int_0^\infty\mathrm d^{|E|}\tau\, e^{-\sum_e q_e \tau_e} \prod_{v \in V} \frac{e^{-p_v\max_{e\ni v}\tau_e}}{p_v}.
 \label{eq:edge-integral}
\end{align}
Remarkably, this gives a dual organization of the same Feynman diagram: the fundamental time variables now reside on edges, while the vertices act as graph local couplings among incident edge times! 

This integral representation, with the ``max"-functions in the exponents, also looks very similar to the ``tropical" representation for canonical forms of graph associahedra. But the attentive reader will notice that \eqref{eq:edge-integral} has one more integration variable than the familiar tropical representations! As shown for ordinary associahedra in \cite{LargeNAmp} and easily generalized to all graph associahedra (in physical kinematics as defined in \cite{Cosmohedron}), a simple manipulation of the tropical representation leads to this simple one-higher-dimensional form.  The cosmological polytope thus discovers conventional time evolution and an ``edge time" evolution description of graph associahedra as literally dual (in the sense of vertex-edge duality) shadows of a single unified object.
\section{Recursion relations}
Splitting the integration domains of the connected part of \eqref{eq:vertex-integral} and the edge representation \eqref{eq:edge-integral} according to the smallest vertex and edge time, respectively, we obtain the familiar recursion relations for the connected contribution and the full wavefunction via the same mechanism. Iterating these recursions generates two complementary tubing pictures.
\\ \\ 
{\bf Smallest vertex --- }We begin with the vertex representation, focusing on the connected contribution, for which we retain the connected piece of each bulk-to-bulk propagator. Stripping the factors of $1/(2y_e)$ for each edge gives:
\begin{equation}
 \mathcal{A}(G)=
 \int_0^\infty
 \prod_{v\in V}\left(d\tau_v\,e^{-x_v\tau_v}\right)
 \prod_{vv'\in E}e^{-y_{vv'}|\tau_v-\tau_{v'}|}.
 \label{eq:amp_int}
\end{equation}
Split the integration domain according to the smallest vertex
time, denoted $\tau_w$. In each region, we make the change of variables
$\tau_v=\tau_w+\tau_v^w$, where $\tau_w^w =0$, which gives:
\begin{align}
\mathcal{A}(G)
={}&
\sum_{w\in V}
\int_0^\infty d\tau_w\,
e^{-\mathcal{E}_{\rm tot}\tau_w}
\nonumber\\
&\times
\int_0^\infty
\prod_{v\neq w}
\left(d\tau_v^w\,e^{-x_v\tau_v^w}\right)
\prod_{uv\in E}
e^{-y_{uv}|\tau_u^w-\tau_v^w|}.
\notag
\end{align}
Integration over the smallest vertex time supplies the total-energy
pole. The second factor is exactly $\mathcal{A}(G\setminus w)$,
with vertex $w$ removed and the energies of its incident edges
added to the external energies of the adjacent vertices. This gives the vertex-deletion recursion for the connected contribution
\begin{align}
\mathcal{A}(G)
=
\frac{1}{\mathcal{E}_{\rm tot}}
\sum_{v \in V}
\mathcal{A}(G \setminus v).
\end{align}
Here $\mathcal{A}$ factorizes over connected components and each
isolated vertex contributes $\mathcal{A}(v)=1/p_v$.

Iterating the recursion produces a sum over maximal tubings:
collections of connected \emph{induced} subgraphs, or tubes,
such that any two tubes are nested or disjoint and no edge joins
two disjoint tubes. For the three-site chain, $C_3$, this gives
\begin{align}
\mathcal{A}({C_3}) &= \frac{1}{\vcenter{\hbox{\includegraphics[scale=0.5]{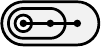}}}}+\frac{1}{\vcenter{\hbox{\includegraphics[scale=0.5]{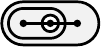}}}}+\frac{1}{\vcenter{\hbox{\includegraphics[scale=0.5]{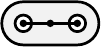}}}}+\frac{1}{\vcenter{\hbox{\includegraphics[scale=0.5]{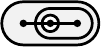}}}}+\frac{1}{\vcenter{\hbox{\includegraphics[scale=0.5]{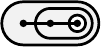}}}}. \notag
\end{align}
The full wavefunction is expanded as the alternating sum 
\begin{equation}
 \Psi_G
 =
 \frac{1}{\prod_e(2y_e)}
 \sum_{F\subseteq E}(-1)^{|F|}
 \mathcal{A}(G\setminus F).
 \label{eq:wavefunction-connected-expansion}
\end{equation}
where we sum over all possible subsets of edges, $F$, of the graph. Here and below, deleting an edge retains its energy as an external energy at each surviving endpoint, so the surviving incidence coordinates are unchanged. For the three-site chain, suppressing the overall factor $1/2y_e$ for each edge, this recursion yields
\begin{align}
\Psi_{C_3} & \propto \frac{1}{\vcenter{\hbox{\includegraphics[scale=0.54]{figs/pt3_t1_a}}}}+\frac{1}{\vcenter{\hbox{\includegraphics[scale=0.54]{figs/pt3_t2_a}}}}+\frac{1}{\vcenter{\hbox{\includegraphics[scale=0.54]{figs/pt3_t5_a}}}}+\frac{1}{\vcenter{\hbox{\includegraphics[scale=0.54]{figs/pt3_t3_a}}}}+\frac{1}{\vcenter{\hbox{\includegraphics[scale=0.54]{figs/pt3_t4_a}}}} \notag \\
&-  \frac{1}{\vcenter{\hbox{\includegraphics[scale=0.55]{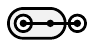}}}}-\frac{1}{\vcenter{\hbox{\includegraphics[scale=0.55]{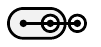}}}}-\frac{1}{\vcenter{\hbox{\includegraphics[scale=0.55]{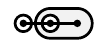}}}}-\frac{1}{\vcenter{\hbox{\includegraphics[scale=0.55]{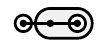}}}}+\frac{1}{\vcenter{\hbox{\includegraphics[scale=0.55]{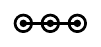}}}}. \notag
\end{align}
In the second line, disjoint tubes may be joined by an edge of the
original graph, since the tubings are defined on the corresponding
edge-deleted subgraphs.
\\ \\
{\bf Smallest edge --- } Applying the same argument to \eqref{eq:edge-integral} gives the \textit{old fashioned perturbation theory} expansion of the wavefunction. In the edge integral, we split the integral into regions according to the smallest edge time, denoted $\tau_f$, and make the change of variables $\tau_{e} =\tau_{f}+\tau_{e}^f$, with $\tau_f^f=0$, which gives
\begin{equation}
\begin{aligned}
\Psi_G
={}&
\sum_{f\in E}
\int_0^\infty d\tau_f\,
e^{-\mathcal{E}_{\rm tot}\tau_f}
\\
&\times
\int_0^\infty
\prod_{e\neq f}
\left(d\tau_e^f\,e^{-q_e\tau_e^f}\right)
\prod_{v\in V}
\frac{e^{-p_v\max_{e\ni v}\tau_e^f}}{p_v}.
\end{aligned} \notag
\end{equation}
Integration over $\tau_f$ again produces the total-energy pole. The remaining integral is the wavefunction of the graph with
edge $f$ removed, taking $\max\varnothing=0$ for isolated vertices. Summing over all edges, $f$, gives the old fashioned perturbation theory recursion \cite{CosmoPolys}
\begin{align}
 \Psi_G = \frac{1}{\mathcal{E}_{\text{tot}}}\sum_{f \in E} \Psi_{G \setminus f }.
\end{align}
Here $\Psi_{G \setminus f}$ factorizes over its connected components, and each isolated vertex contributes $\Psi_v= 1/p_v$. 

Iterating this recursion gives the original tubing representation
of the wavefunction, again as a sum over maximal tubings. Here tubes are connected subgraphs that need not be induced; any two tubes are nested or disjoint, but disjoint tubes may now be joined by an edge of the original graph. For the four-site chain, $C_4$, this gives
\begin{align}
\tilde\Psi_{C_4} = \frac{1}{\vcenter{\hbox{\includegraphics[scale=0.47]{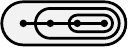}}}}+\frac{1}{\vcenter{\hbox{\includegraphics[scale=0.47]{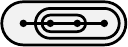}}}}+\frac{1}{\vcenter{\hbox{\includegraphics[scale=0.47]{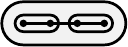}}}}+\frac{1}{\vcenter{\hbox{\includegraphics[scale=0.47]{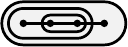}}}}+\frac{1}{\vcenter{\hbox{\includegraphics[scale=0.47]{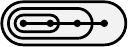}}}}, \notag
\end{align}
where $\tilde\Psi_G=(\prod_{v\in V}p_v)\Psi_G$ denotes the wavefunction stripped of the $1/p_v$ poles associated with the individual vertices.
\section{Outlook}

In this letter we have shown how the cosmological polytope is the parent of two complementary notions of ``time'' underlying the wavefunction, both the conventional ``vertex-centered" time of Feynman diagrams and its dual ``edge-time". We end by discussing a few obvious avenues of research opened up by this simple realization. 
\\ \\
\textbf{Vertex--edge duality for hypergraphs} ---
The two time representations suggest asking how the wavefunction
changes when we exchange vertices and edges in the graph itself.
In the incidence dual, each edge of the original graph becomes a
vertex, and each original vertex becomes a hyperedge containing the
new vertices corresponding to its incident edges. A vertex of degree
greater than two thus gives rise to a hyperedge with more than two
vertices. Since the inequalities defining the dual cone depend only
on the incidence relations, they provide a natural starting point
for extending cosmological polytopes to hypergraphs and studying
how their canonical functions transform under this duality. Some combinatorial aspects of this duality will be explored in \cite{cosmOrder}.
\\ \\
\textbf{Large $n$ wavefunctions} --- The naive computation of wavefunction coefficients, even for a single graph with a large $n$ number of vertices gives us a sum over many terms whose number is exponentially large in $n$, much like tree amplitudes for Tr$(\phi^3)$ theory. But in \cite{LargeNAmp}, it has been realized that there is a dual representation for amplitudes that exposes an amazing simplicity when $n$ is large. Indeed, for the case of a line graph, where the graph associahedron is the associahedron itself, one of the representations of \cite{LargeNAmp} is precisely the edge-time representation we have given here, and indeed the results of \cite{LargeNAmp} were a motivation for looking for such a representation for general graphs entering the wavefunction. We can thus use exactly the same ideas of \cite{LargeNAmp} to expose a striking simplicity in the large $n$ behavior of wavefunctions for graphs $G_n$. For example, if we focus on $n$-site chain graphs, $C_n$, using the methods from \cite{LargeNAmp} applied to the time-integrals defining $\Psi_{C_n}$, we find that in the simple regime where all $x_v =x$ and $y_e=y$, the leading large $n$ behavior is simply: 
\begin{equation*}
    \Psi_{C_n} \xrightarrow[n\to \infty]{} \left[ \frac{2}{j_{\frac{2y}{x},1}\, x}\right]^{2n} \, .
 \end{equation*}
where $j_{\frac{2y}{x},1}$ stands for the location of the first zero of the Bessel function $J_{2y/x}$. If instead we had $x_v \neq x$ and $y_e \neq y$, under the assumptions of a smooth variation for $(x_v,y_e)$ along the chain, from \cite{LargeNAmp} the expression above just turns into a product over the $n$ vertices.  The ideas from \cite{LargeNAmp} can also be generalized beyond flat space by applying the frequency integrals that recover the wavefunction for conformally coupled scalars in general FRW backgrounds, providing a route to analogous large-$n$ results in cosmology.
\\ \\
\textbf{Kinematic flow from geometry} --- In FRW backgrounds, the resulting cosmological integrals satisfy a system of differential equations, known as kinematic flow \cite{KinFlow}. The upcoming work \cite{KinFLowFromPolys} shows that the Laplace transform with added twists,
$$
\Psi_G^{(\boldsymbol\alpha)}=
 \mathcal{N}
 \int_{C_G^\vee} \prod_{v} \frac{d \eta_v \, e^{- \eta_v x_v}}{\eta_v^{1+\alpha_v}} \prod_e d \eta_e \, e^{-\eta_e y_e}
$$
(with $\alpha_v$ being the twists) naturally selects certain faces of the cosmological polytope as base functions which close under action of the kinematic derivative. Furthermore, the entire structure of the matrix of the differential equations (and thus the kinematic flow) has some avatar in the cosmological polytope and its dual: in the dual picture faces corresponding to basis functions couple only to boundary faces, and no face is part of the basis functions if it lies on the plane spanned by the twists. This extends the remarkable point made in this letter, that the cosmological polytope not only encodes different bulk-time evolution avatars but, given the twist measure, it also encodes the integrated FRW wavefunction by almost trivially fixing the kinematic flow structure.
\\ \\
\textbf{Cosmological Correlators} --- Another interesting direction is to extend these ideas for equal-time correlators. There have been different proposals on connecting correlators with positive geometries \cite{GeometryCosmoCorr,Figueiredo:2025daa,Glew:2026von}. But to mirror the ideas followed in this letter, the most useful one is to start from the representation in \cite{Arkani-Hamed:2025mce}, which gives the graph-correlator as a time-integral of the type \eqref{eq:vertex-integral} but integrated over the full time domain, and with the bulk-to-bulk propagators replaced by Feynman propagators, $G_{BB}^F$, and the bulk-to-boundary propagators also replaced by Feynman propagators, $e^{-x_v \eta_v}\to e^{-x_v |\eta_v|}$. Finally we can remove all the absolute values in the exponents by uplifting the graph-correlator integral to:
\begin{equation*}
\Omega_G =
\int_{\mathcal{K}_G}
d^{V}\eta_v
d^{n}t_a
d^{E}\eta_e\,
e^{-\sum_v x_v \eta_v
-\sum_a E_a t_a
-\sum_e y_e \eta_e},
\end{equation*}
where $n$ is the number of external legs in the graph, and $E_a$ is the energy of the external leg $a$. Then, the integration domain is:
\begin{equation*}
\mathcal{K}_G =
\left\{
(\eta_v,t_a,\eta_e)\,\middle|\,
\begin{aligned}
t_a &\geq 0,\\
t_a+2\eta_{v(a)} &\geq 0,\\
\eta_e+\eta_v-\eta_w &\geq 0,\\
\eta_e-\eta_v+\eta_w &\geq 0
\end{aligned}
\right\},
\quad e=(v,w).
\end{equation*}
Interpreting $\Omega_G$ as a Laplace transform, we learn that the correlator can be obtained by restricting the canonical form, $\Omega_G$, to the sub-space $x_v=\sum_{a\ni v} E_a$ -- giving us yet another interpretation of the correlator in terms of geometry.
\\ \\
\textbf{Connection to Cosmohedra} --- In this letter, we have seen how, for a single graph, the different formulations for $\Psi_G$ -- via cosmological polytopes, graph associahedra and time-integrals -- have been unified in a simple way. An obvious question is how does this generalize to the full wavefunction $\Psi_n$, which we obtain by summing over all graphs. So far, we only know of a single geometric object which ``glues'' all graph-associahedra together in a non-trivial way -- the cosmohedron \cite{Cosmohedron} -- correctly capturing the combinatorics of the full wavefunction. However, the canonical form of this object does not give directly the wavefunction -- and therefore connecting it to the standard time-integral picture seems out of reach. But, motivated by our findings in this letter, one might wonder if we should instead start by looking for an integral representation for the full wavefunction -- like the ones we have for the amplitudes via tropical integrals -- and from it discover a larger object which, just like the cosmological polytope, unites cosmohedra with both the standard time-evolution from Feynman diagrams and it's edge-time dual.

\section*{Acknowledgements}
N.A.H. is supported by the DOE (Grant No. DE-SC0009988), the Simons Collaboration on Celestial Holography, the ERC UNIVERSE+ synergy grant, and the Carl B. Feinberg cross-disciplinary program in innovation at the IAS. C.F. is supported by the Society of Fellows at Harvard University. R.G. and F.V. are supported by the Institute for Advanced Study and an anonymous donor.

\bibliography{Refs}

\end{document}